\documentclass[sigconf,authorversion]{acmart}

\usepackage{booktabs} 

\usepackage{soul}
\usepackage{url}
\usepackage{datetime2}

\usepackage{hyperref}

\usepackage{acronym}
\usepackage{relsize}
\makeatletter
\AtBeginDocument{%
  \renewcommand*{\AC@hyperlink}[2]{%
    \begingroup
      \hypersetup{hidelinks}%
      \hyperlink{#1}{#2}%
    \endgroup
  }%
}
\makeatother

\usepackage{cleveref}

\usepackage{balance}

\newcommand{\bigon}[1]{$\mathcal{O}(N^{#1})$}

\newcommand{\intelreg}{Intel\textregistered}

\newcommand{\inteloptanereg}{{\intelreg} Optane\textsuperscript{TM}}

\newcommand{\intelopafreg}{{\intelreg} Omni-Path Fabric}

\acrodef{fma}[FMA]{fused multiply-add}
\acrodef{mkl}[MKL]{Intel Math Kernel Library}
\acrodef{fft}[FFT]{Fast Fourier Transform}
\acrodef{imdt}[IMDT]{Intel Memory Drive Technology}
\acrodef{dram}[DRAM]{dynamic random access memory}
\acrodef{ram}[RAM]{random access memory}
\acrodef{nvm}[NVM]{non-volatile memory}
\acrodef{nvme}[NVMe]{\acs{nvm} Express}
\acrodef{hdd}[HDD]{Hard Disk Drive}
\acrodef{ssd}[SSD]{Solid-State Drive}
\acrodef{numa}[NUMA]{non-uniform memory access}
\acrodef{os}[OS]{operating system}
\acrodef{sku}[SKU]{stock keeping unit}
\acrodef{sdm}[SDM]{software-defined memory}
\acrodef{scm}[SCM]{storage-class memory}
\acrodef{sdm-f}[SDM-F]{\acs{sdm} for Fabrics}
\acrodef{sdm-s}[SDM-S]{\acl{sdm} for \ac{scm}}
\acrodef{vm}[VM]{virtual memory}
\acrodef{cpu}[CPU]{central processing unit}
\acrodef{pci}[PCI]{Peripheral Component Interconnect}
\acrodef{pcie}[PCIe]{\acl{pci} Express}
\acrodef{gemm}[GEMM]{GEneral Matrix Multiplication}
\acrodef{blas}[BLAS]{Basic Linear Algebra Subprograms}
\acrodef{eri}[ERI]{electron-repulsion integral}
\acrodef{flop}[FLOP]{floating-point operation}
\acrodef{flops}[FLOPS]{floating-point operations per second}
\acrodef{hf}[HF]{Hartree-Fock method}
\acrodef{qpi}[QPI]{Intel QuickPath Interconnect}
\acrodef{hpc}[HPC]{high performance computing}

\acmYear{2018} 

\copyrightyear{2018} 
\setcopyright{licensedusgovmixed}

\acmDOI{10.1145/3286475.3286479}

\acmISBN{978-1-4503-6113-2/18/11}
\acmPrice{15.00}

\acmConference[MCHPC'18]{MCHPC'18: Workshop on Memory Centric High Performance Computing}{November 11, 2018}{Dallas, TX, USA}
\acmBooktitle{MCHPC'18: Workshop on Memory Centric High Performance Computing (MCHPC'18), November 11, 2018, Dallas, TX, USA}

\begin{document}
\title{Evaluation of Intel Memory Drive Technology Performance for Scientific Applications}

\author{Vladimir Mironov}
\orcid{0000-0002-9454-5823}
\affiliation{
	\institution{Department of Chemistry,\\Lomonosov Moscow State University}
    \streetaddress{Leninskie Gory 1/3}
    \city{Moscow}
    \postcode{119991}
    \country{Russian Federation}
} \email{vmironov@lcc.chem.msu.ru}

\author{Andrey Kudryavtsev}
\affiliation{%
  \institution{Intel Corporation}
  \streetaddress{1900 Prairie City rd}
  \city{Folsom}
  \state{California}
  \postcode{95630}
  \country{USA}
} \email{andrey.o.kudryavtsev@intel.com}

\author{Yuri Alexeev}
\affiliation{
	\institution{Argonne National Laboratory,\\Computational Science Division}
    \city{Argonne}
    \state{Illinois}
    \postcode{60439}
    \country{USA}
} \email{yuri@alcf.anl.gov}

\author{Alexander Moskovsky}
\affiliation{
	\institution{RSC Technologies}
    \city{Moscow}
    \country{Russian Federation}
} \email{moskov@rsc-tech.ru}

\author{Igor Kulikov}
\affiliation{
	\institution{Institute of Computational Mathematics and Mathematical Geophysics SB RAS}
    \city{Novosibirsk}
    \country{Russian Federation}
} \email{kulikov@ssd.sscc.ru}

\author{Igor Chernykh}
\affiliation{
	\institution{Institute of Computational Mathematics and Mathematical Geophysics SB RAS}
    \city{Novosibirsk}
    \country{Russian Federation}
} \email{chernykh@ssd.sscc.ru}

\renewcommand{\shortauthors}{V. Mironov et al.}

\begin{abstract}
In this paper, we present benchmark data for \ac{imdt}, which is a new generation of Software-defined Memory (SDM) based on Intel ScaleMP collaboration and using 3D XPoint \textsuperscript{TM} based Intel \acp{ssd} called Optane. We studied \ac{imdt} performance for synthetic benchmarks, scientific kernels, and applications. We chose these benchmarks to represent different  patterns for computation and accessing data on disks and memory. To put performance of \ac{imdt} in comparison, we used two memory configurations: hybrid \ac{imdt} DDR4/Optane and DDR4 only systems. The performance was measured as a percentage of used memory and analyzed in detail. We found that for some applications DDR4/Optane hybrid configuration outperforms DDR4 setup by up to 20\%.
\end{abstract}

%
%

\ccsdesc[500]{Hardware~Non-volatile memory}
\ccsdesc[300]{Computing methodologies~Massively parallel and high-performance simulations}
\ccsdesc[300]{Applied computing}
\ccsdesc[300]{Software and its engineering~Memory management}

\keywords{Intel Memory Drive Technology, Solid State Drives, Intel Optane, ScaleMP}

\maketitle

\acresetall

\section{Introduction}

In the recent years the capacity of system memory for high performance computing (HPC) systems has not been kept with the pace of the increased \ac{cpu} power. The amount of system memory often limits the size of problems that can be solved.
System memory is typically based on \ac{dram}. \ac{dram} prices have significantly grown up in the recent year. In 2017, \ac{dram} prices were growing up approximately 10-20\% quarterly~\cite{trendforce}. As a result, memory can contribute up to 90\% to the cost of the servers.

A modern memory system is a hierarchy of storage devices with different capacities, costs, latencies, and bandwidths intended to reduce price of the system. It makes a perfect sense to introduce yet another level in the memory hierarchy between \ac{dram} and hard disks to drive price of the system down. \acp{ssd} are a good candidate because they are cheaper than \ac{dram} up to 10 times. What is more important, over the last 10 years, \acp{ssd} based on NAND technology emerged with higher read/write speed and Input/Ouput Operations per Second (IOPS) metric than hard disks.

Recently, Intel announced~\cite{optaneannounce} a new \ac{ssd} product based on novel 3D XPoint\textsuperscript{TM} technology under the name \inteloptanereg. It was developed to overcome the drawbacks of NAND-technology: block-based memory addressing and limited write endurance. To be more specific, with 3D XPoint each memory cell can be addressed individually and write endurance of 3D XPoint memory is significantly higher than NAND \acp{ssd}. As a result, 3D XPoint flash memory can be used instead of \ac{dram}, albeit as a slow memory, which can be still an attractive solution given that Intel Optane is notably cheaper than \ac{ram} per gigabyte. A novel \ac{imdt} allows to use Intel Optane drives as a system memory. Another important advantage of 3D XPoint compared to \ac{dram} is that it has a high density of memory cells, which allows to build compact systems with massive memory banks.

\begin{figure*}[!t]
		\centering
		\includegraphics[width=0.65\linewidth]{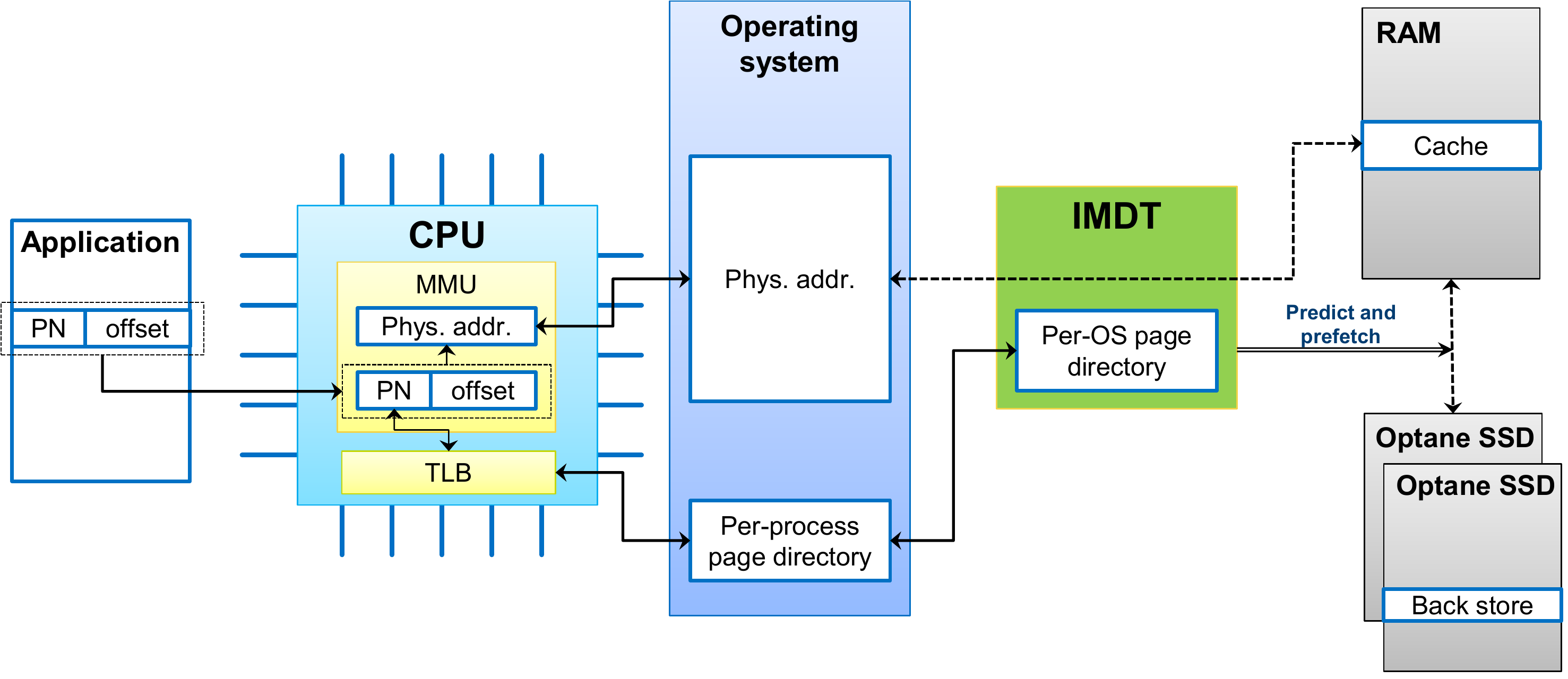}
		\caption{This figure describes how \acl{imdt} works. Solid lines represent inquiry, dashed lines represent data transfer, and double lines represent commands issued.}
		\label{fig:imdt}
\end{figure*}

In this work, we evaluated the capabilities of Intel Optane drives together with \ac{imdt} for numerical simulations requiring large amount of memory. We started with the overview of \ac{imdt} technology in section 2. In section 3, we described the methodology. In Sections 4 and 5 we described all benchmarks and corresponding performance results. In section 6 we discussed the performance results, and in Section 7 we presented our conclusions and plans for the future work.

\section{Overview of Intel Memory Drive Technology}

For effective use of Intel Optane in hybrid \ac{ram}-\ac{ssd} memory systems, Intel corporation and ScaleMP developed a technology called \ac{imdt} \cite{imdtweb,imdtweb2}. \ac{imdt} integrates the Intel Optane into the memory subsystem and makes it appear like \ac{ram} to the \acl{os} and applications. Moreover, \ac{imdt} increases memory capacity beyond \ac{ram} limitations and performs in a completely transparent manner without any changes in \acl{os} and applications. The key feature of \ac{imdt} is that \ac{ram} is effectively used as cache. As a result, \ac{imdt} can achieve good performance compared to all-\ac{ram} systems for some applications at a fraction of the cost as we have shown in this paper.

ScaleMP initially has developed a technology to make virtual \ac{numa} system using high speed node interconnect of modern high performance computational clusters. \ac{numa} systems are typically defined as any contemporary multi-socket system. It allows a processor to access memory at varying degrees of latency or ``distance'' (e.g. memory attached to another processor), over a network or fabric.
In some cases, this fabric is purpose-built for such processor communication, like \intelreg\ QuickPath and UltraPath Interconnects (\intelreg\ QPI and UPI respectively). In other cases, standard fabrics such as \ac{pcie} or \intelopafreg\ are used for the same purpose along with \ac{sdm} to provide memory coherency, operating as if additional memory was installed in the system.

Accessing memory at varying lower performance over networks has proven to be feasible and useful by using predictive memory access technologies that support advanced caching and replication, effectively trading latency for bandwidth. This is exactly what \ac{imdt} is doing to enable \ac{nvm} to be used as system memory. Instead of doing it over fabric, however, it does so with storage. With \ac{imdt}, most of the Intel Optane capacity is transparently used as an extension to the \ac{dram} capacity of the system.

\ac{imdt} is implemented as an \ac{os}-transparent virtual machine (\Cref{fig:imdt}). In \ac{imdt}, Intel Optane \acp{ssd} are used as part of the system memory to present the aggregated capacity of the \ac{dram} and \ac{nvm} installed in the system as one coherent, shared memory address space. No changes are required to the operating system, applications, or any other system components. Additionally \ac{imdt} implements advanced memory access prediction algorithms to optimize memory access performance.

A popular approach to virtualize disk memory is to store part of \ac{vm} pages on special disk partition or file is implemented in all popular operating systems nowadays. However, the resulting performance is very sensitive not only to the storage speed but also to \ac{vm} manager implementation. It is very important to correctly predict which memory page on disk will be needed soon and to load it in \ac{ram} to avoid program spinning in a page fault state. The built-in \ac{os} swap in Linux kernel is not very intelligent and usually affected by this problem. On the contrary, \ac{imdt} analyzes memory access patterns and prefetches the data into the \ac{ram} ``cache'' (\Cref{fig:imdt}) before it is used, resulting in better performance.

\ac{imdt} leverage the low-latency media access provide by Intel Optane \acp{ssd}. NAND \ac{ssd} latency cannot be improved by simply aggregating multiple drives. Transitioning to Intel Optane \acp{ssd} is another step forward to the reductions of the gap between \ac{dram} and \ac{ssd} performance by using lower latency media based on the 3D XPoint technology. However, \ac{dram} still has lower latency than Intel Optane, which can potentially affect the performance of applications with DRAM+Optane configuration studied in this paper.

\section{Methodology}

\ac{imdt} architecture is based on the hypervisor layer which manages paging exclusively. This makes a hybrid memory transparent from one side, however, standard \ac{cpu} counters become unavailable to performance profiling tools. Thus, we took an approach to make a comparison with \ac{dram}-based system side by side. The efficiency metric was calculated as a ratio of software defined performance counters, if available, or simply the ratio of the time-to-solution on \ac{dram}-based system and \ac{imdt}-based system.

\subsection{Hardware and software configuration}
In this study, we used dual-socket Intel Broadwell (Xeon E5 2699 v4, 22 cores, 2.2 GHz) node with latest version of BIOS. We have used two memory configurations for this node. In the first configuration, it was equipped with 256 GB DDR4 registered ECC memory (16$\times$16~GB Kingston 2133~MHz DDR4) and four \inteloptanereg\ SSDs P4800X (320~GB memory mode). We used Intel Memory Drive Technology 8.2 to expand system memory with Intel \acp{ssd} up to approximately 1,500 GB. In the second configuration, the node was exclusively equipped by 1,536 GB of DDR4 registered ECC memory (24$\times$64~GB Micron 2666~MHz DDR4). In both configurations we used a stripe of four 400 GB Intel DC P3700 SSD drives as a local storage.  Intel Parallel Studio XE 2017 (update 4) was used to compile the code for all benchmarks. Hardware counters on non-IMDT setup were collected using \intelreg\ Performance Counter Monitor~\cite{pcmweb}.

\subsection{Data size representation}
\ac{imdt} assumes that all data is loaded in the \ac{ram} before it is actually used. It is important to note that if the whole dataset fits in the \ac{ram}, it is very unlikely that it will be moved to the Optane disks. In this case, the difference between \ac{imdt} and \ac{ram} should be negligible. The difference will be more visible only when the data size is significantly larger than the available \ac{ram}. Since the performance results are connected to the actual \ac{ram} size, we find more convenient to represent benchmark sizes in parts of \ac{ram} in \ac{imdt} configuration (256 GB) and not in GB or problem dimensions. Such representation of data sets is more general and the results can be extrapolated to different hardware configurations.

\section{Description of Benchmarks}

In this section, we described various types of benchmarks to evaluate performance of \ac{imdt}. We broadly divided benchmarks in three classes -- synthetic benchmarks, scientific kernels, and scientific applications. The goal is to test performance for a diverse set of scientific applications, which have different memory access patterns with various memory bandwidth and latency requirements.

\subsection{Synthetic benchmarks}

\subsubsection{STREAM} \cite{McCalpin1995} is a simple benchmark commonly used to measure sustainable bandwidth of the system memory and corresponding computation rate for a few simple vector kernels. In this work, we have used multi-threaded implementation of this benchmark. We studied memory bandwidth for a test requiring $\approx500$~GB memory allocation on 22, 44, and 88 threads.

\subsubsection{Polynomial benchmark}\label{sssec:poly_desc}
was used to compute polynomials of various degree of complexity. Polynomials are commonly used in mathematical libraries for fast and precise evaluation of various special functions. Thus, they are virtually present in all scientific programs. In our tests, we calculated polynomials of predefined degree over a large array of double precision data stored in memory.

Memory access pattern is similar to the STREAM benchmark. The only difference that we can finely tune the arithmetic intensity of the benchmark by changing the degree of computed polynomials. From this point of view STREAM benchmark is a particular case of the polynomial benchmark when the polynomial degree is zero (\textit{STREAM copy}) or one (\textit{STREAM scale}). We used Horner's method of polynomial evaluation which is efficiently translated to the \ac{fma} operations.

We have calculated performance for polynomials of degrees 16, 64, 128, 256, 512, 768, 1024, 2048, and 8192 using various data sizes (from 50 to 900 GB). We studied two data access patterns. In the first one we just read the value from the array of arguments, calculate the polynomial value and add it to a thread-local variable. There is only one (\textit{read}) data stream to the \ac{imdt} disk storage in this case. In another case the result of polynomial calculation updates corresponding value in the array of arguments. There are two data streams here (\textit{read} and \textit{write}). Arithmetic intensity of this benchmark was calculated as follows:
\begin{equation}
\label{eqn:ai}
AI = 2\cdot \frac{polynomial\ degree}{sizeof(double)},
\end{equation}
where factor two corresponds to the one addition and one multiplication for each polynomial degree in Horner's method of polynomial evaluation.

\subsubsection{\acsu{gemm}}
(\acl{gemm}) is one of the core routines in \ac{blas} library. It is a level 3 \ac{blas} operation defining matrix-matrix operation. \ac{gemm} is often used for performance evaluations and it is our first benchmark to evaluate \ac{imdt} performance. \ac{gemm} is a compute-bound operation with \bigon{3} arithmetic operations and \bigon{2} memory operations, where $N$ is a leading dimension of matrices. Arithmetic intensity grows as \bigon{} depending on matrix size and is flexible. The source code of the benchmark used in our tests is available here \cite{gemmgithub}.

\subsection{Scientific kernels}

\subsubsection{LU decomposition}
(where ``LU'' stands for ``lower upper'' of a matrix, and also called LU factorization) is a commonly used kernel in a number of important linear algebraic problems like solving system of linear equations, finding eigenvalues, etc. In current study, we used \ac{mkl}~\cite{mklweb} implementations of LU decomposition, more specifically \texttt{dgetrf} and \texttt{mkl\_dgetrfnpi}. We also studied the performance of an LU decomposition algorithm using tile algorithm, which dramatically improved performance of \ac{imdt}. The source code of the latter was taken from the \textit{hetero-streams} code base \cite{heterostreams}.

\subsubsection{\acf{fft}} is an algorithm that samples a signal over a period of time or space and divides it into its frequency components. \ac{fft} is an important kernel in many scientific codes. In this work, we have studied the performance of the \ac{fft} implemented in \ac{mkl} library \cite{mklweb}. We have used three-dimensional decomposition of the $N \times N \times N$ grid data. The benchmark sizes were $N = (500\div5800)$ resulting in 0.001-1.5 TB memory footprint.

\subsection{Scientific applications}

\subsubsection{LAMMPS} 
(Large-scale Atomic/Molecular Massively Parallel Simulator) is a popular molecular simulation package developed in Sandia National Laboratory \cite{Plimpton1995}. Its main focus is a force-field based molecular dynamics. We have used scaled Rhodopsin benchmark distributed with the source code. Benchmark set was generated from the original chemical system (32,000 atoms) by its periodical replication in $X, Y$ (8 times) and $Z$ (8-160 times) dimensions. The largest chemical system comprises 328,000,000 atoms ($\approx1.1$ TB memory footprint). Performance metrics -- number of molecular dynamics steps per second.

\subsubsection{GAMESS}
(General Atomic and Molecular Electronic Structure System) is one of the most popular quantum chemistry packages. It is a general purpose program, where a large number of quantum chemistry methods are implemented. We used the latest version of code distributed from GAMESS website \cite{gamesswebsite}. In this work, we have studied the performance of the Hartree-Fock method. We have used stacks of benzene molecules as a model chemical system. By changing the number of benzene molecules in stack we can vary memory footprint of the application. 6-31G(d) basis set was used in the simulations.

\subsubsection{AstroPhi} is a hyperbolic PDE engine which is used for numerical simulation of astrophysical problems \cite{AstroPhiGitHub}. AstroPhi realizing a multi-component hydrodynamic model for astrophysical objects interaction. The numerical method of solving hydrodynamic equations is based on a combination of an operator splitting approach, Godunov's method with modification of Roe's averaging, and a piecewise-parabolic
method on a local stencil \cite{Popov2007,Popov2008}. The redefined system of equations is used to guarantee the non-decrease of entropy \cite{Godunov2014} and for speed corrections \cite{illposed}. The detailed description of a numerical method can be found in \cite{KULIKOV201571}. In this work, we used the numerical simulation of gas expansion into vacuum for benchmarking. We have used 3D arrays with up to $2000^3$ size ($\approx1.5$ TB memory footprint) for this benchmark.

\subsubsection{PARDISO} is a package for sparse linear algebra calculations. It is a part of Intel MKL library \cite{mklweb}. In our work, we studied the performance of the Cholesky decomposition of sparse (\bigon{} non-zero elements) $N \times N$ matrices, where $N=(5,10,20,25,30,35,40)\cdot10^6$. Memory footprint of benchmarks varied from 36 to 790 GB.

\subsubsection{Intel-QS} (former qHiPSTER) is a distributed high-performance implementation of a quantum simulator on a classical computer, that can simulate general single-qubit gates and two-qubit controlled gates \cite{qs-arxiv}. The code is fully parallelized with MPI and OpenMP. The code is architectured in the way that memory consumption exponentially grows as more qubits are being simulated. We benchmarked a provided quantum FFT test for 30-35 qubit simulations. 35 qubits simulation required more than 1.5TB of memory. The code used in our benchmarks was taken from Intel-QS repository on Github \cite{qsgithub}.

\section{Results}

\subsection{Synthetic benchmarks}

\begin{figure}[t]
	\centering
	\includegraphics[width=\linewidth]{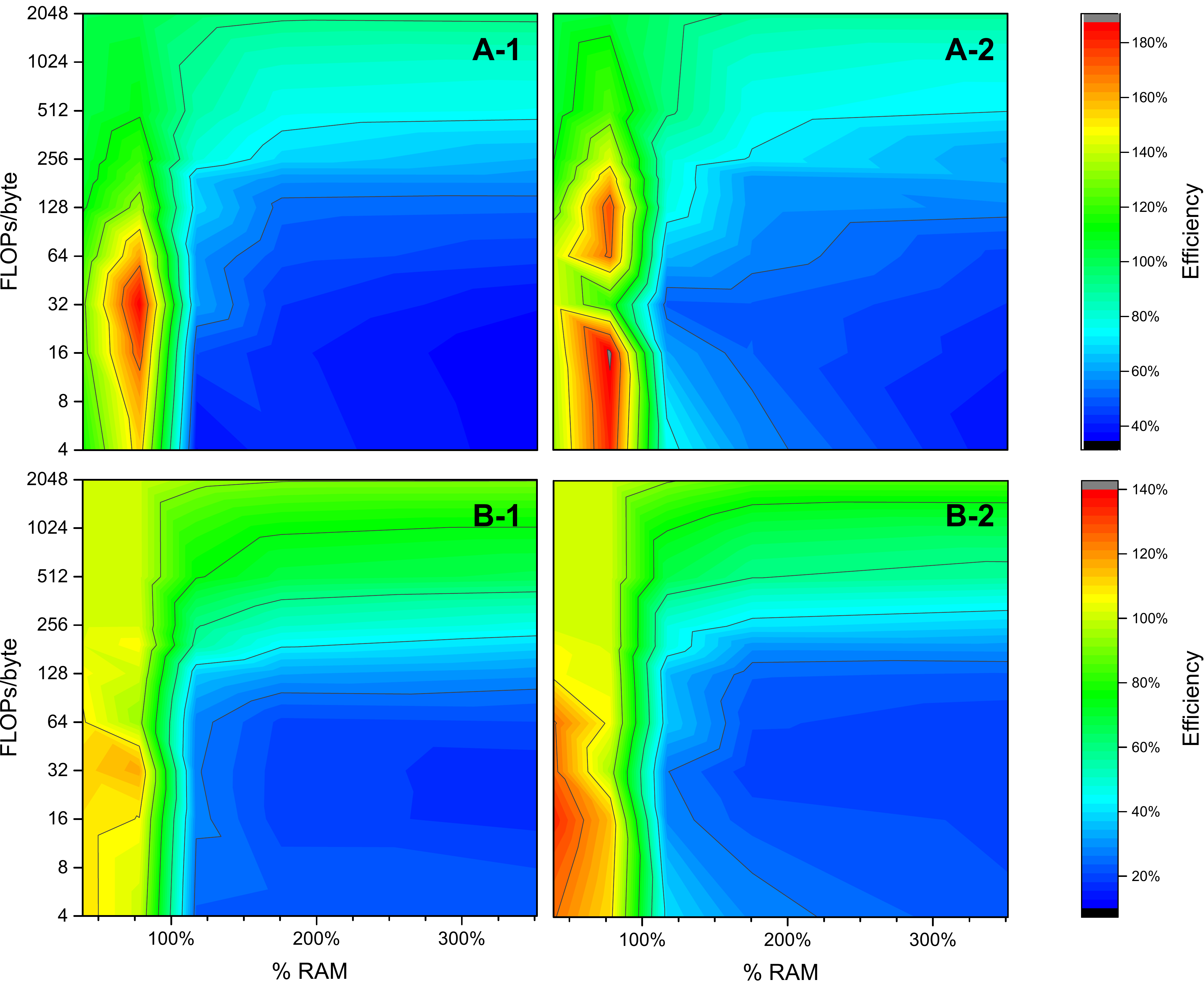}
	\caption{Polynomial benchmark results. (A) -- one data stream for 44 (A-1) and 88 (A-2) threads. (B) -- two data streams for 44 (B-1) and 88 (B-2) threads. The efficiency is denoted by color with a legend on corresponding row. See text for more details.}
	\label{fig:polyresults}
\end{figure}

\subsubsection{STREAM} benchmark was used as a reference to a worst case scenario, where application has low CPU utilization and high memory bandwidth requirements. 
We obtained 80~GB/s memory bandwidth for the \ac{dram}-configured node, while for \ac{imdt}-configured node we got 10~GB/s memory bandwidth for the benchmarks requesting maximum available memory. In other words, we are comparing best case scenario for \ac{dram} bandwidth with the worst case scenario on \ac{imdt}. Thus, we can expect the worst possible efficiency of $10/80 = 12.5\%$ \ac{imdt} vs \ac{dram}. It should be noted that running benchmarks which fit in \ac{dram} cache of \ac{imdt} results in the bandwidth equal to (80-100~GB/s), which is comparable to the \ac{dram} bandwidth. This is what we expected and it is the proof that \ac{imdt} utilizes optimally \ac{dram} cache. The measured bandwidth actually depends only on the number of threads and it was higher for the less concurrent jobs. It applies only to \ac{imdt} benchmarks requesting memory smaller than the size of \ac{dram} cache. Memory bandwidth of \ac{dram}-configured node does not depend on the workload size nor on the number of threads.

\subsubsection{Polynomial benchmark.}\label{sssec:poly_res}
Results of the polynomial benchmarks are presented in \Cref{fig:polyresults}. As one can see in \Cref{fig:polyresults}, patterns of efficiency are very similar. If the data fits in the \ac{ram} cache of \ac{imdt} then \ac{imdt} typically shows better performance than \ac{dram}-configured node, especially for short polynomials. High concurrency (88 threads, \Cref{fig:polyresults} (A-2 and B-2)) is also beneficial to \ac{imdt} in these benchmarks. However, a better efficiency can be obtained for benchmarks with higher order of polynomials. In terms of arithmetic intensity (\cref{eqn:ai}), it is required to have at least 256 \acp{flop} per byte to get \ac{imdt} efficiency close to $100\%$. It will be discussed later in detail (see \Cref{ssec:perf_analysis}).

\subsubsection{GEMM benchmark.} 
\begin{figure}[t]
	\centering
	\includegraphics[width=0.9\linewidth]{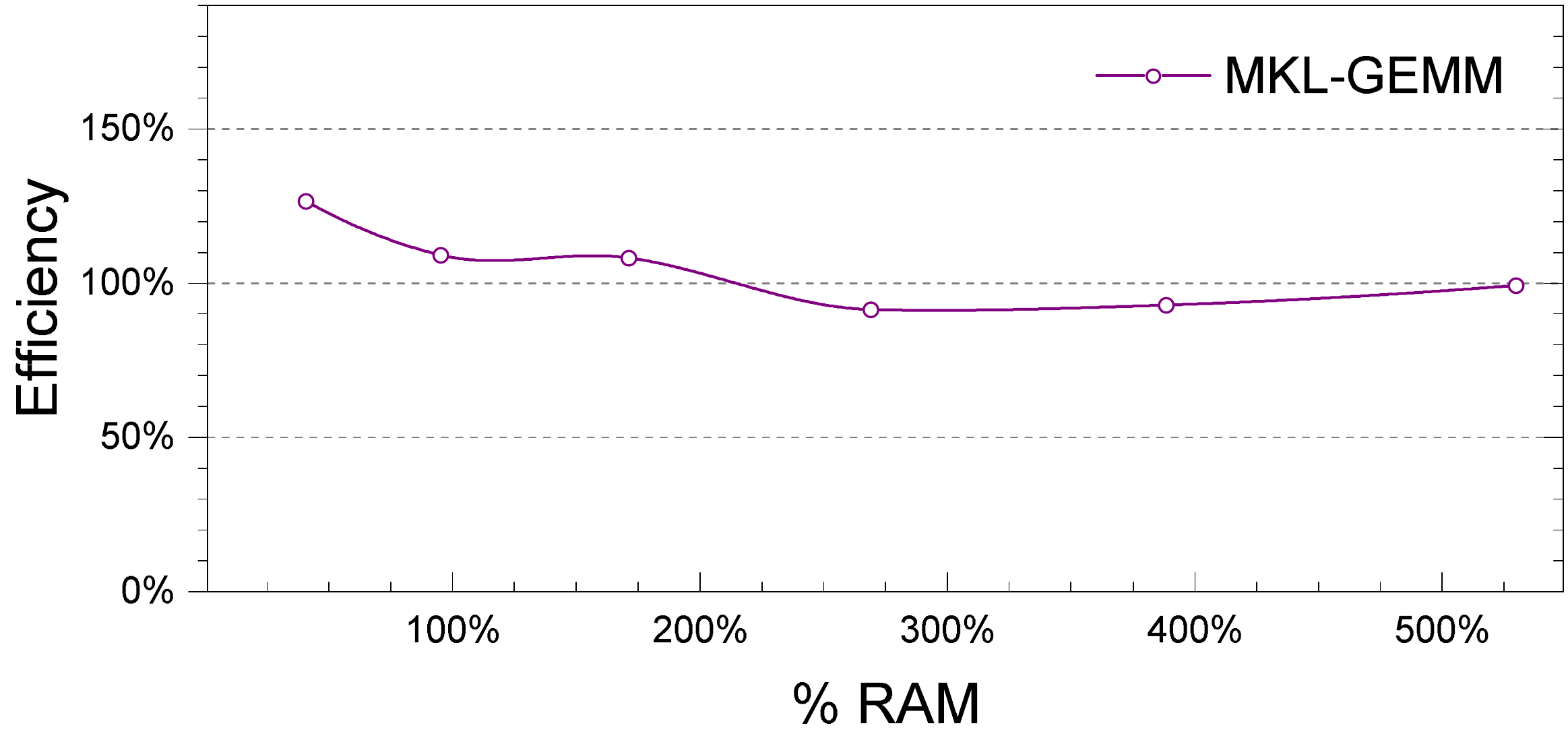}
	\caption{IMDT efficiency plot for GEMM benchmark. Higher efficiency is better. 100\% efficiency corresponds to DRAM performance.}
	\label{fig:gemm}
\end{figure}

According to our benchmarks shown in \Cref{fig:gemm}, GEMM shows very good efficiency for every problem size. All observed efficiencies vary from 90\% for large benchmarks to 125\% for small benchmarks. Such efficiency is expected because GEMM is purely compute bound. To be more specific, the arithmetic intensity even for a relatively small GEMM benchmark is much higher than the required value of 250 FLOP/byte per data stream, which was estimated in polynomial benchmarks. In our tests, we have used custom (``segmented'') GEMM benchmark with improved data locality: all matrices are stored in a tiled format (arrays of ``segments'') and matrix multiplication goes tile-by-tile. Thus the arithmetic intensity of all benchmarks is constant and equal to the arithmetic intensity of a simple tile-by-tile matrix multiplication. It is approximately equal to 2 FLOPs multiplied by tile dimension and divided by the size of data type in bytes (4 for \texttt{float} and 8 for \texttt{double}). In our benchmark with a single-precision GEMM with typical tile dimension size of $\approx$43000, the arithmetic intensity is $\approx$21500 FLOPs/byte, which is far beyond the required 250 FLOPs/byte.

\begin{figure}[t]
	\centering
	\includegraphics[width=0.9\linewidth]{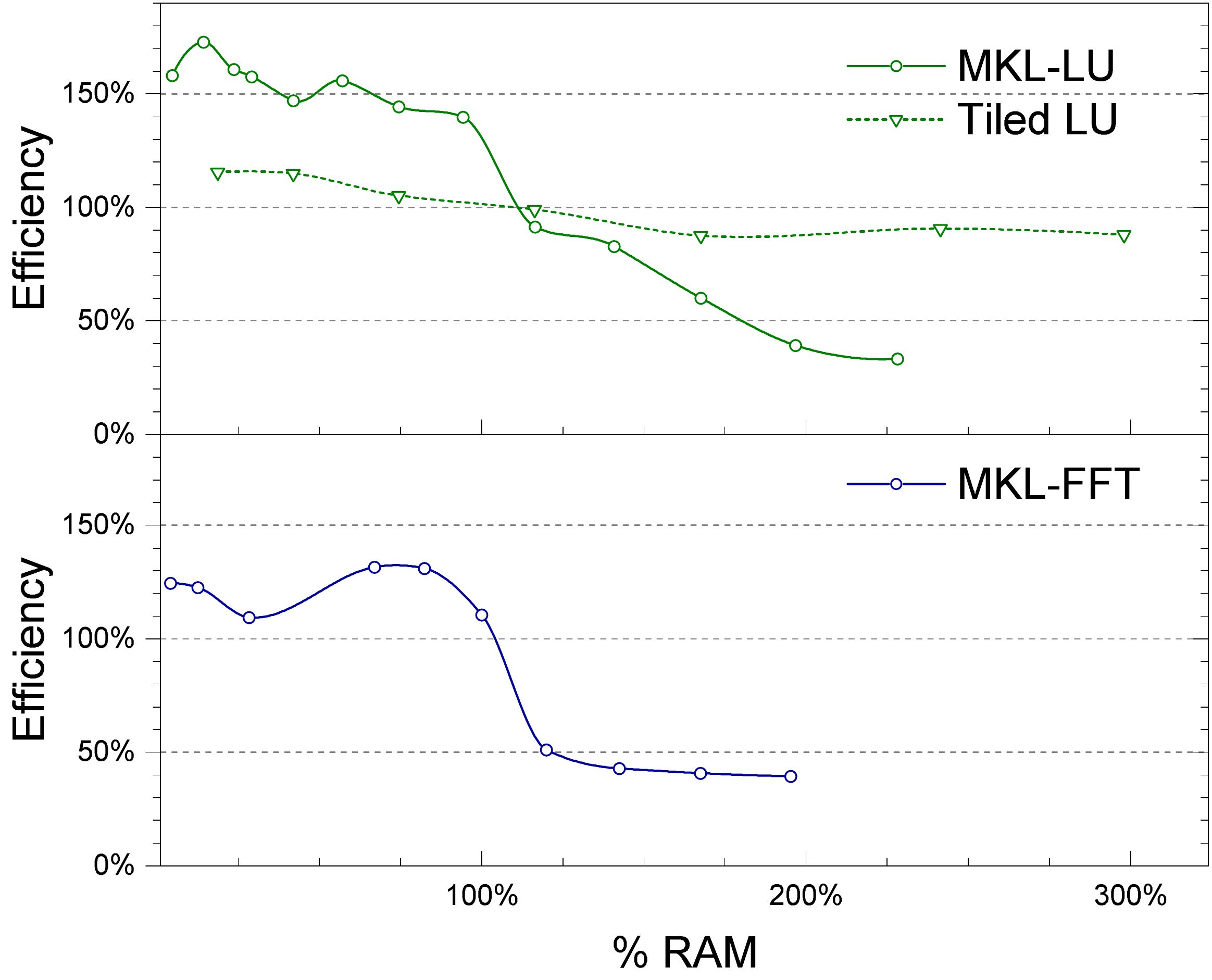}
	\caption{IMDT efficiency plots for LU and \ac{fft} benchmarks. Two implementations (MKL and tiled) of LU decomposition were benchmarked. Higher efficiency is better. 100\% efficiency corresponds to DRAM performance.}
	\label{fig:mklresults}
\end{figure}

\subsection{Scientific kernels}

\subsubsection{LU decomposition} The efficiency of LU decomposition implemented in \ac{mkl} library strongly depends on the problem size. The efficiency is excellent when a matrix fits into the memory (\Cref{fig:mklresults}). We observe about $150\%-180\%$ speedup on \ac{imdt} for small matrices. However, the efficiency decreases down to $\approx30\%$ for very large matrices with leading dimension equal or greater than $\approx2\cdot10^5$. This result was unexpected; in fact, our LU implementation calls \ac{blas} level 3 functions such as GEMM, which has excellent efficiency on \ac{imdt} as we demonstrated in previous section. We can provide two explanations for unfavorable memory access patterns in LU decomposition. First one is the partial pivoting which interchanges rows and columns in the original matrix. Second, matrix is stored as a contiguous array in memory that is known for its inefficient memory access to the elements of the neighboring columns (rows in case of Fortran). Both problems are absent in a special tile-based LU decomposition benchmark implemented in \textit{hetero-streams} code base. We also ran benchmarks for this optimized LU decomposition benchmark. Tiling of the matrix not only improved the performance by about 20\% for both ``\ac{dram}'' and ``\ac{imdt}'' memory configurations, but also improved the efficiency of \ac{imdt} to $\approx$90\% (see \Cref{fig:mklresults}). Removing of pivoting only without introducing matrix tiling does not significantly improve the efficiency of LU decomposition.

\subsubsection{Fast Fourier Transform.} The results of \ac{mkl}-\ac{fft} benchmark are similar to those obtained for \ac{mkl}-LU as shown in \Cref{fig:mklresults}. For small problem sizes the efficiency of \ac{imdt} exceeds $100\%$, but for large benchmarks the efficiency drops down to $\approx40\%$. Performance drop occurs at $100\%$ of \ac{ram} utilization. \ac{fft} problems typically have relatively small arithmetic intensity (small ratio of \acp{flop}/byte). Thus, obtaining relatively low \ac{imdt} efficiency was expected. We still believe that the \ac{fft} benchmark can be optimized for memory locality to improve \ac{imdt} efficiency even higher (see
\cite{Bailey:1989:FEH:76263.76288,cormen1998fftooc,Akin:2016:FNM:2985294.2985306}
for the examples of memory-optimized \ac{fft} implementations).

\subsection{Scientific applications}
\begin{figure}[t]
	\centering
	\includegraphics[width=\linewidth]{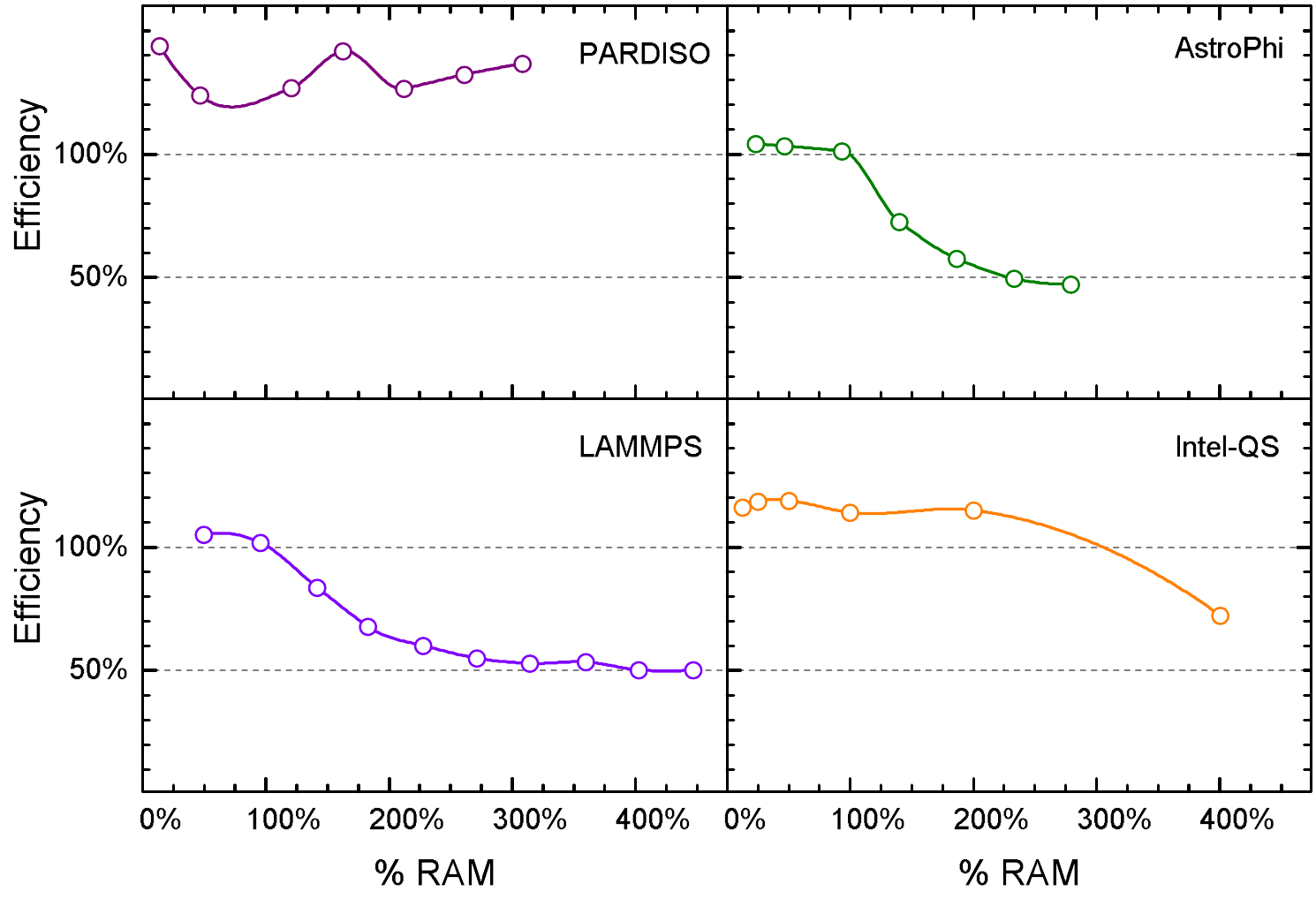}
	\caption{IMDT efficiency plots for various scientific applications. Higher efficiency is better. 100\% efficiency corresponds to DRAM performance.}
	\label{fig:apps}
\end{figure}

The benchmarking results for different scientific applications are shown in \Cref{fig:apps} and \Cref{fig:hfresults}. The applications are PARDISO, AstroPhi, LAMMPS, Intel-QS, and GAMESS. All applications except PARDISO show similar efficiency trends. When a benchmark requests memory smaller than the amount of available \ac{dram}, the application performance on the \ac{imdt}-configured node is typically higher than for \ac{dram}-configured node. At a certain threshold, which is typically a multiple of \ac{dram} size, the \ac{imdt} efficiency declines based on the CPU Flop/S and memory bandwidth requirements.

\subsubsection{MKL-PARDISO}
PARDISO is very different from other studied benchmarks. The observed IMDT efficiency is 120-140\% of ``\ac{dram}''-configured node for all studied problem sizes. It was not very surprising because Cholesky decomposition is known to be compute intensive. MKL-PARDISO is optimized for the disk-based out-of-core calculations resulting in excellent memory locality to access data structures. As a result, this benchmark always benefits from faster access to the non-local \ac{numa} memory on \ac{imdt} which results in the improved performance on ``\ac{imdt}''-configured node.

\subsubsection{AstroPhi}
In \Cref{fig:apps} we presented the efficiency plot of Lagrangian step of the gas dynamic simulation, which is the most time consuming step ($>90\%$ compute time). This step describes the convective transport of the gas quantities with the scheme velocity for the gas expansion into vacuum problem. The number of \acp{flop}/byte is not very high and the efficiency plot follows the trend we described above. We observe a slow decrease in efficiency down to $\approx$50\% when the data does not fit into \ac{dram} cache of IMDT. Otherwise the efficiency is close to 100\%.

\begin{figure}[t]
	\centering
	\includegraphics[width=0.85\linewidth]{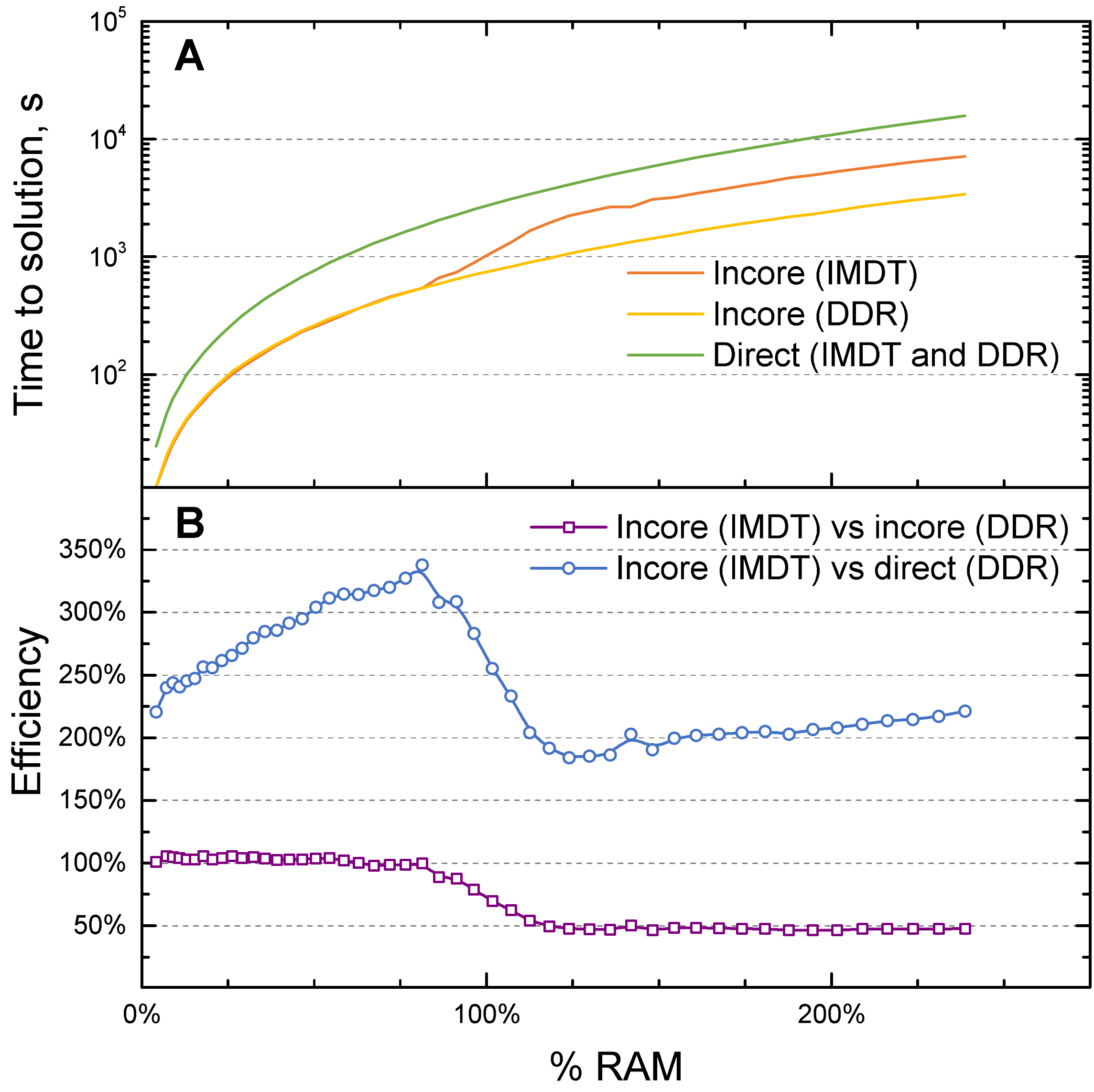}
	\caption{GAMESS Hartree-Fock simulation (10 iterations) for stacks of benzene molecules with 6-31G(d) basis set. (A) Time to solution in seconds, lower is better. (B) \ac{imdt} efficiency, higher efficiency is better, 100\% efficiency corresponds to \ac{dram} performance. The performance of the direct Hartree-Fock on ``\ac{imdt}'' and ``\ac{dram}''-configured node (see text for details) is the same ((A), green line).}
	\label{fig:hfresults}
\end{figure}

\subsubsection{LAMMPS}
We studied the performance of the molecular dynamics of the Rhodopsin benchmark provided with LAMMPS distribution. It is an all-atom simulation of the solvated lipid bilayer surrounding Rhodopsin protein. The calculations are dominated by the long-range electrostatics interactions in particle-particle mesh algorithm. The results of benchmarks are presented in \Cref{fig:apps}. It is obvious that LAMMPS efficiency follows the same pattern as AstroPhi and \ac{fft}: it is more than 100\% when tests fit in the \ac{dram} cache of the \ac{imdt} and it is dropping down when tests do not fit. For tests with high memory usage the efficiency is $\approx$50\%.

\subsubsection{Intel-QS}

We benchmarked Intel-QS using provided quantum FFT example for 30-35 qubits. Actually, each additional qubit doubles the amount of memory required for the job. For that reason we had to stop at 35 qubit test which occupy about 1~TB of memory. The observed IMDT efficiency was greater than 100\% for 30-34 qubits and drops down to $\approx$70\% at 35 qubit simulation. A significant portion of the simulation take \ac{fft} steps. Thus, degradation of the performance at high memory utilization was not surprising. However, the overall efficiency is almost two times better than for \ac{fft} benchmark.

\subsubsection{GAMESS}
We studied the performance of the two \ac{hf} algorithms. \ac{hf} is solved iteratively and for each iteration a large number of \acp{eri} need to be re-computed (direct \ac{hf}) or read from disk or memory (conventional \ac{hf}).
In the special case of conventional \ac{hf} called incore \ac{hf}, \acp{eri} are computed once before \ac{hf} iterations and stored in \ac{dram}. In the subsequent \ac{hf} iterations, the computed \acp{eri} are read from memory. We benchmarked both direct and incore \ac{hf} methods. The former algorithm has small memory footprint, but re-computation of all \acp{eri} each iteration (typical number of iterations is 20) results in much longer time to solution compared to incore \ac{hf} method if \acp{eri} fit in memory.

The performance of the direct \ac{hf} method on ``\ac{dram}'' and ``\ac{imdt}''-configured nodes is very similar (see \Cref{fig:hfresults} (A), green line). However, the performance of the incore method differs between ``\ac{dram}'' and ``\ac{imdt}'' (see \Cref{fig:hfresults} (A), red and yellow lines). The efficiency shown in \Cref{fig:hfresults} (B) for incore \ac{imdt} vs incore \ac{dram} (purple line) behaves similar to other benchmarks -- when benchmarks fits in the \ac{dram} cache of the \ac{imdt} then the efficiency is close to 100\%, otherwise it decreases to $\approx$50\%. But for incore \ac{imdt} vs direct \ac{dram} (blue line) the efficiency is much better. The efficiency varies between approximately 200\% and 350\%. Thus, \ac{imdt} can be used to speed up Hartree-Fock calculations when the amount of \ac{dram} is not available to fit all \acp{eri} in memory.

\begin{figure}[t]
	\centering
	\includegraphics[width=0.9\linewidth]{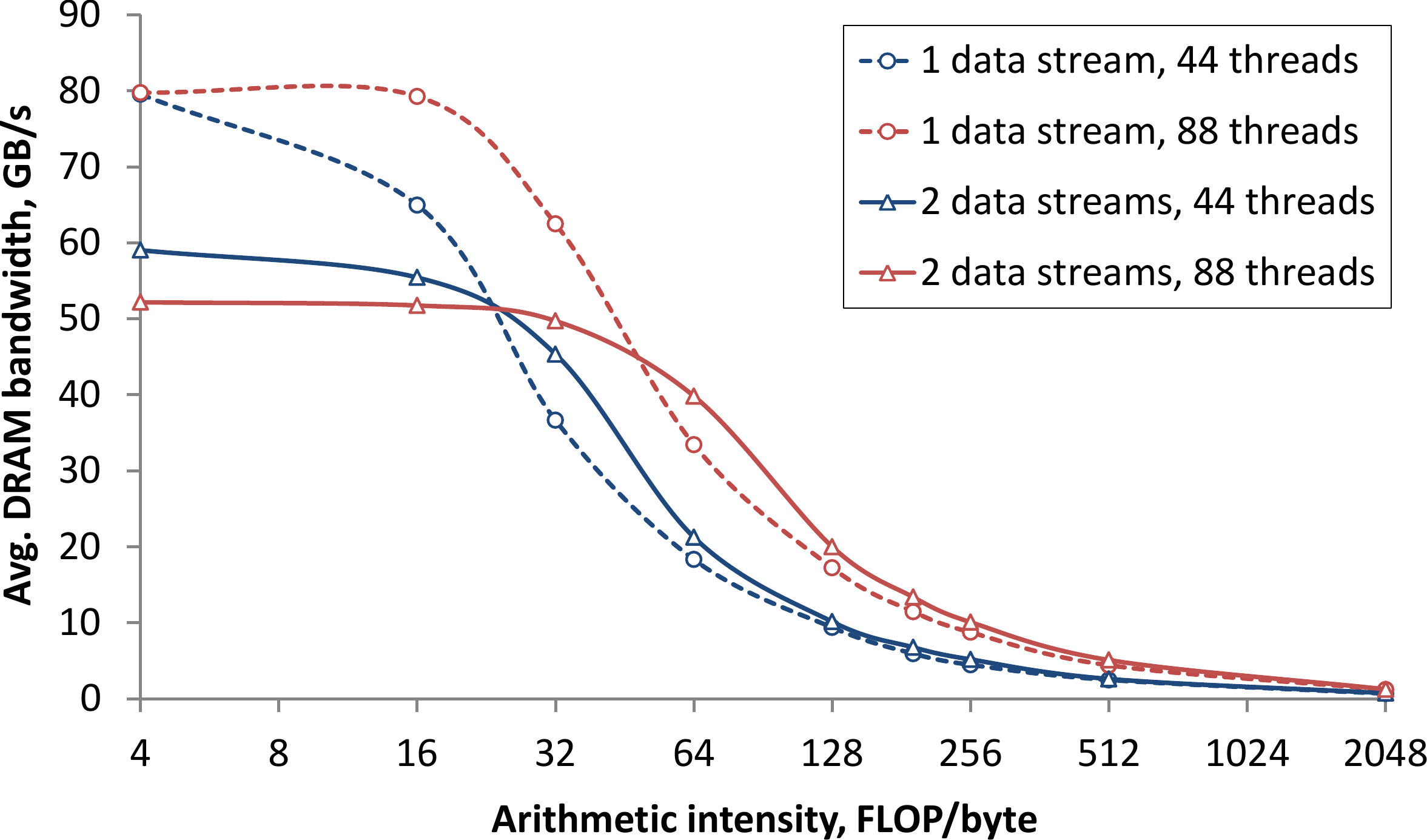}
	\caption{Average DRAM bandwidth in the polynomial benchmark depending on the arithmetic intensity (polynomial degree), number of data streams (1~data stream -- read-only access pattern, 2 data streams -- read/write access pattern, see text for details), and the number of working threads. The results were obtained by Intel Processor Counter Monitor (sum of System READ and WRITE counters).}
	\label{fig:poly-bandwidth}
\end{figure}

\subsection{Analysis of IMDT performance}\label{ssec:perf_analysis}
Modern processors can overlap data transfer and computation very efficiently. A good representative example is the Polynomial benchmark (see \Cref{sssec:poly_desc}). When the polynomial degree is low the time required to move data from system memory to CPU is much higher than the time of polynomial computation (\Cref{sssec:poly_res}, \Cref{fig:polyresults}). In this case, the performance is bound by memory bandwidth. By increasing the amount of computation, the overlap between data transfer and computation becomes more efficient and the benchmark gradually transforms from a memory-bound to a compute-bound problem. Increasing of arithmetic intensity is achieved by increasing the degree of polynomials (see \cref{eqn:ai}).

On \Cref{fig:poly-bandwidth} the dependence of average DRAM bandwidths on the arithmetic intensity is shown for polynomial benchmarks with different number of data streams. The improvement of overlap between data transfer and computation is observed at about 16-32~FLOP/byte. The computation of low-order polynomials (less than 16~FLOP/byte) is not limited by the compute power, resulting in high memory bandwidth. The bandwidth value depends on I/O pattern (i.e. the number of data streams) and it is limited by the DRAM memory bandwidth, which is about 80~GB/s.

The bandwidth dependence on the number of data streams of low-order polynomials results from NUMA memory access. If the benchmark was optimized for NUMA, the highest bandwidth for one and two data streams would be the same. However, in IMDT architecture, while application thread accesses the remote NUMA node for writes, IMDT places the data to the DRAM attached to the local NUMA node. This can significantly reduce the pressure on the cross-socket link and as a result the performance can become better than DRAM based system performance. It is exactly what we observed in our tests when all the data fits in the DRAM cache (see \Cref{fig:polyresults}).

When the arithmetic intensity grows beyond 16~FLOP/byte, the memory bandwidth starts decreasing. At 64~FLOP/byte and beyond the benchmark becomes compute bound. It means that the memory bandwidth does not depend on the number of data streams but on the availability of computational resources (i.e. number of threads). However, the memory bandwidth decreases slowly with the arithmetic intensity (\Cref{fig:poly-bandwidth}). Taking into  account that the memory bandwidth of our \ac{imdt} system is capped by 10~GB/s, we expect that only those benchmarks that are below this threshold will have good efficiency. It is expected that it will apply to all benchmarks with different problem sizes. In terms of the arithmetic intensity, it corresponds to $\approx128-256$~FLOP/byte. This correlation is shown on \Cref{fig:polyresults}. The same analysis can be applied to any benchmark to estimate the potential efficiency of the \ac{imdt} approach.

\subsection{Summary}

To sum up our benchmarking results, our tests show that there is virtually no difference between using \ac{dram} and \ac{imdt} if a benchmark requires memory less than the amount of available \ac{ram}. \ac{imdt} correctly handles these cases, if the test fits in \ac{ram} and there is no need to use Optane memory. In fact, \ac{imdt} frequently outperforms \ac{ram} because \ac{imdt} has advanced memory management system. The situation is very different for large tests. For some tests like dense linear algebra, PARDISO and Intel-QS efficiency remains high, while for other applications like LAMMPS, AstroPhi, and GAMESS 
the efficiency slowly declines to about 50\%. Even in the latter case IMDT can be attractive for scientific users since it enables larger problem sizes to be addressed.

\section{Discussion}

One of the most important benefits of \ac{imdt} is that it significantly reduces data traffic trough \ac{qpi} bus on \ac{numa} systems. For example, GEMM unoptimized benchmark on ``\ac{dram}''-configured node performs about 20-50\% slower for large datasets compared to a small ones. The main reason is overloaded \ac{qpi} bus. When a benchmark saturates \ac{qpi} bandwidth then it causes CPU stalls waiting for data. \ac{qpi} bandwidth in our system is 9.6~GT/s (1~GT/s$=10^9$ transfers per second) or $\approx$10~GB/s unidirectional ($\approx$20~GB/s bidirectional) and it can easily become a bottleneck. It is an inherent issue of multisocket \ac{numa}-systems which adversely affects performance of not only GEMM, but any other applications.

There are a few ways to resolve this issue. For example, in optimized GEMM implementation~\cite{gemmgithub} matrices are split to tiles, which are placed in memory intelligently taking into the account the ``first-touch'' memory allocation policy in Linux OS. As a result, \ac{qpi} bus load drops to 5-10\% and performance significantly improves achieving almost theoretical peak. It was observed in our experiments with GEMM by using Intel Performance Counter Monitor (PCM) software~\cite{pcmweb}.

There is no such issue with \ac{imdt} and performance is consistently close to theoretical peak even for the unoptimized GEMM implementation. \ac{imdt} provides optimal access to the data on the remote \ac{numa} node improving the efficiency of almost all applications. This is why we almost never seen in practice a very low \ac{imdt} efficiency even for strongly memory-bandwidth bound benchmarks like \ac{fft} and AstroPhi. Theoretical efficiency minimum of 12.5\% was observed only for the specially designed synthetic benchmarks like STREAM and polynomial benchmark.

However, \ac{imdt} is not a solution to all memory-related issues. For example, it cannot help in situations when an application has random memory access patterns across a large number of memory pages with a low degree of application parallelism. While the performance penalty is not very high for DRAM memory, frequent access of the IMDT backstore on \ac{ssd} can be limited by the bandwidth of Intel Optane \ac{ssd}, and IMDT can only compensate for that if the workload has a high degree of parallel memory accesses (using many threads or many processes concurrently).
In such cases, it may be beneficial to redesign data layout for better locality of data structures. In this work, we observed it when we ran \ac{mkl} implementation of LU decomposition. Switching to the tiled implementation of the LU algorithm results in the significantly improved efficiency of \ac{imdt} because of better data locality. The similar approach can be applied to other applications. However, it is beyond the scope of this paper and it is a subject for our future studies.

\section{Conclusions and future work}

\ac{imdt} is a revolutionary technology that flattens the last levels of memory hierarchy: \ac{dram} and disks.
One of the major \ac{imdt} advantages is the high density of memory. It will be feasible in the near future to build systems with many terabytes of Optane memory. In fact, the bottleneck will not be the amount of Optane memory, but the amount of available \ac{dram} cache for \ac{imdt}. It is currently possible to build an \ac{imdt} system with 24~TB of addressable memory (with 3~TB DRAM cache), which is not possible to build with \ac{dram}. Even if it was possible to build a such system, \ac{imdt} offers a more cost effective solution.

There are HPC applications with large memory requirements. It is a common practice for such applications to store data in parallel network storage or use distributed memory. In this case, the application performance can be limited by network bandwidth. There is now another alternative which is to use DRAM+Optane configuration with \ac{imdt}. In theory, the bandwidth of the multiple striped Optane drives exceeds network bandwidth. \ac{imdt} especially benefits the applications that poorly scale on the multi-node environment due to high communication overhead. A good example of such application is a quantum simulator. Indeed, Intel-QS simulator efficiency shown in \Cref{fig:apps} is excellent compared to other applications. Another good application that fits profile is the visualization of massive amount of data. We plan to explore the potential of \ac{imdt} for such applications in our future work.

IMDT prefetching subsystem analyzes memory access patterns in the real time and makes appropriate adjustments according to workload characteristics. This feature is crucial for IMDT performance and differentiates it from other solutions such as OS swap. We plan to analyze it in detail in our future work.

This work is important because we systematically studied performance of \ac{imdt} technology for a diverse set of scientific applications. We have demonstrated that applications and benchmarks exhibit reasonable performance level, when the system main memory is extended with the help of \ac{imdt} by Optane \ac{ssd}. In some cases, we have seen DRAM+Optane configuration to outperform DRAM-only system by up to 20\%. Based on performance analysis, we provide recipes how to unlock full potential of \ac{imdt} technology. It is our hope that this work will educate professionals about this new exciting technology and promote its wide-spread use.

\section{Acknowledgements}
This research used the resources of the Argonne Leadership Computing Facility, which is a U.S. Department of Energy (DOE) Office of Science User Facility supported under Contract DE-AC02-06CH11357. We gratefully acknowledge the computing resources provided and operated by the Joint Laboratory for System Evaluation (JLSE) at Argonne National Laboratory. We thank the \intelreg\ Parallel Computing Centers program for funding, ScaleMP team for technical support, RSC Group and Siberian Supercomputer Center ICMMG SB RAS for providing access to hardware, and Gennady Fedorov for help with \intelreg\ PARDISO benchmark. This work was partially supported by the Russian Fund of Basic Researches grant 18-07-00757, 18-01-00166 and by the Grant of the Russian Science Foundation (project 18-11-00044).

\bibliographystyle{ACM-Reference-Format}
\bibliography{imdt}


\begin{thebibliography}{22}


\ifx \showCODEN    \undefined \def \showCODEN     #1{\unskip}     \fi
\ifx \showDOI      \undefined \def \showDOI       #1{#1}\fi
\ifx \showISBNx    \undefined \def \showISBNx     #1{\unskip}     \fi
\ifx \showISBNxiii \undefined \def \showISBNxiii  #1{\unskip}     \fi
\ifx \showISSN     \undefined \def \showISSN      #1{\unskip}     \fi
\ifx \showLCCN     \undefined \def \showLCCN      #1{\unskip}     \fi
\ifx \shownote     \undefined \def \shownote      #1{#1}          \fi
\ifx \showarticletitle \undefined \def \showarticletitle #1{#1}   \fi
\ifx \showURL      \undefined \def \showURL       {\relax}        \fi
\providecommand\bibfield[2]{#2}
\providecommand\bibinfo[2]{#2}
\providecommand\natexlab[1]{#1}
\providecommand\showeprint[2][]{arXiv:#2}

\bibitem[\protect\citeauthoryear{??}{Ast}{2018}]%
        {AstroPhiGitHub}
 \bibinfo{year}{2018}\natexlab{}.
\newblock \bibinfo{title}{{AstroPhi. The hyperbolic PDE engine }}.
\newblock
\newblock
\urldef\tempurl%
\url{https://github.com/IgorKulikov/AstroPhi}
\showURL{%
\tempurl}


\bibitem[\protect\citeauthoryear{??}{het}{2018}]%
        {heterostreams}
 \bibinfo{year}{2018}\natexlab{}.
\newblock \bibinfo{title}{{Hetero Streams Library}}.
\newblock
\newblock
\urldef\tempurl%
\url{https://github.com/01org/hetero-streams}
\showURL{%
\tempurl}


\bibitem[\protect\citeauthoryear{??}{mkl}{2018}]%
        {mklweb}
 \bibinfo{year}{2018}\natexlab{}.
\newblock \bibinfo{title}{{Intel Math Kernel Library}}.
\newblock
\newblock
\urldef\tempurl%
\url{https://software.intel.com/mkl}
\showURL{%
\tempurl}


\bibitem[\protect\citeauthoryear{??}{imd}{2018a}]%
        {imdtweb}
 \bibinfo{year}{2018}\natexlab{a}.
\newblock \bibinfo{title}{{Intel Memory Drive Technology}}.
\newblock
\newblock
\urldef\tempurl%
\url{https://www.intel.com/content/www/us/en/software/intel-memory-drive-technology.html}
\showURL{%
\tempurl}


\bibitem[\protect\citeauthoryear{??}{pcm}{2018}]%
        {pcmweb}
 \bibinfo{year}{2018}\natexlab{}.
\newblock \bibinfo{title}{{Intel Performance Counter Monitor -- A better way to
  measure CPU utilization}}.
\newblock
\newblock
\urldef\tempurl%
\url{www.intel.com/software/pcm}
\showURL{%
\tempurl}


\bibitem[\protect\citeauthoryear{??}{gem}{2018}]%
        {gemmgithub}
 \bibinfo{year}{2018}\natexlab{}.
\newblock \bibinfo{title}{{Segmented SGEMM benchmark for large memory
  systems}}.
\newblock
\newblock
\urldef\tempurl%
\url{https://github.com/ScaleMP/SEG_SGEMM}
\showURL{%
\tempurl}


\bibitem[\protect\citeauthoryear{??}{imd}{2018b}]%
        {imdtweb2}
 \bibinfo{year}{2018}\natexlab{b}.
\newblock \bibinfo{title}{{Software Defined Memory at a Fraction of the DRAM
  Cost -- white paper}}.
\newblock
\newblock
\urldef\tempurl%
\url{https://www.intel.com/content/www/us/en/solid-state-drives/intel-ssd-software-defined-memory-with-vm.html}
\showURL{%
\tempurl}


\bibitem[\protect\citeauthoryear{??}{gam}{2018}]%
        {gamesswebsite}
 \bibinfo{year}{2018}\natexlab{}.
\newblock \bibinfo{title}{{The General Atomic and Molecular Electronic
  Structure System (GAMESS)}}.
\newblock
\newblock
\urldef\tempurl%
\url{http://www.msg.ameslab.gov/gamess/index.html}
\showURL{%
\tempurl}


\bibitem[\protect\citeauthoryear{??}{qsg}{2018}]%
        {qsgithub}
 \bibinfo{year}{2018}\natexlab{}.
\newblock \bibinfo{title}{{The Intel Quantum Simulator}}.
\newblock
\newblock
\urldef\tempurl%
\url{https://github.com/intel/Intel-QS}
\showURL{%
\tempurl}


\bibitem[\protect\citeauthoryear{??}{tre}{2018}]%
        {trendforce}
 \bibinfo{year}{2018}\natexlab{}.
\newblock \bibinfo{title}{{TrendForce}}.
\newblock
\newblock
\urldef\tempurl%
\url{http://www.trendforce.com}
\showURL{%
\tempurl}


\bibitem[\protect\citeauthoryear{A.~Vshivkov, G.~Lazareva, V.~Snytnikov,
  Kulikov, and V.~Tutukov}{A.~Vshivkov et~al\mbox{.}}{2011}]%
        {illposed}
\bibfield{author}{\bibinfo{person}{Vitaly A.~Vshivkov}, \bibinfo{person}{Galina
  G.~Lazareva}, \bibinfo{person}{Alexei V.~Snytnikov}, \bibinfo{person}{I
  Kulikov}, {and} \bibinfo{person}{Alexander V.~Tutukov}.}
  \bibinfo{year}{2011}\natexlab{}.
\newblock \showarticletitle{Computational methods for ill-posed problems of
  gravitational gasodynamics}.
\newblock   \bibinfo{volume}{19} (\bibinfo{date}{05} \bibinfo{year}{2011}).
\newblock


\bibitem[\protect\citeauthoryear{Akin, Franchetti, and Hoe}{Akin
  et~al\mbox{.}}{2016}]%
        {Akin:2016:FNM:2985294.2985306}
\bibfield{author}{\bibinfo{person}{Berkin Akin}, \bibinfo{person}{Franz
  Franchetti}, {and} \bibinfo{person}{James~C. Hoe}.}
  \bibinfo{year}{2016}\natexlab{}.
\newblock \showarticletitle{{FFTs} with Near-Optimal Memory Access Through
  Block Data Layouts: Algorithm, Architecture and Design Automation}.
\newblock \bibinfo{journal}{\emph{J. Signal Process. Syst.}}
  \bibinfo{volume}{85}, \bibinfo{number}{1} (\bibinfo{date}{Oct.}
  \bibinfo{year}{2016}), \bibinfo{pages}{67--82}.
\newblock
\showISSN{1939-8018}
\urldef\tempurl%
\url{https://doi.org/10.1007/s11265-015-1018-0}
\showDOI{\tempurl}


\bibitem[\protect\citeauthoryear{Bailey}{Bailey}{1989}]%
        {Bailey:1989:FEH:76263.76288}
\bibfield{author}{\bibinfo{person}{D.~H. Bailey}.}
  \bibinfo{year}{1989}\natexlab{}.
\newblock \showarticletitle{{FFTs} in External of Hierarchical Memory}. In
  \bibinfo{booktitle}{\emph{Proceedings of the 1989 ACM/IEEE Conference on
  Supercomputing}} \emph{(\bibinfo{series}{Supercomputing '89})}.
  \bibinfo{publisher}{ACM}, \bibinfo{address}{New York, NY, USA},
  \bibinfo{pages}{234--242}.
\newblock
\showISBNx{0-89791-341-8}
\urldef\tempurl%
\url{https://doi.org/10.1145/76263.76288}
\showDOI{\tempurl}


\bibitem[\protect\citeauthoryear{Cormen and Nicol}{Cormen and Nicol}{1998}]%
        {cormen1998fftooc}
\bibfield{author}{\bibinfo{person}{Thomas~H Cormen} {and}
  \bibinfo{person}{David~M Nicol}.} \bibinfo{year}{1998}\natexlab{}.
\newblock \showarticletitle{Performing out-of-core {FFTs} on parallel disk
  systems}.
\newblock \bibinfo{journal}{\emph{Parallel Comput.}} \bibinfo{volume}{24},
  \bibinfo{number}{1} (\bibinfo{year}{1998}), \bibinfo{pages}{5--20}.
\newblock


\bibitem[\protect\citeauthoryear{Godunov and Kulikov}{Godunov and
  Kulikov}{2014}]%
        {Godunov2014}
\bibfield{author}{\bibinfo{person}{S.~K. Godunov} {and} \bibinfo{person}{I.~M.
  Kulikov}.} \bibinfo{year}{2014}\natexlab{}.
\newblock \showarticletitle{Computation of discontinuous solutions of fluid
  dynamics equations with entropy nondecrease guarantee}.
\newblock \bibinfo{journal}{\emph{Computational Mathematics and Mathematical
  Physics}} \bibinfo{volume}{54}, \bibinfo{number}{6} (\bibinfo{date}{01 Jun}
  \bibinfo{year}{2014}), \bibinfo{pages}{1012--1024}.
\newblock
\showISSN{1555-6662}
\urldef\tempurl%
\url{https://doi.org/10.1134/S0965542514060086}
\showDOI{\tempurl}


\bibitem[\protect\citeauthoryear{Kulikov, Chernykh, Snytnikov, Glinskiy, and
  Tutukov}{Kulikov et~al\mbox{.}}{2015}]%
        {KULIKOV201571}
\bibfield{author}{\bibinfo{person}{I.M. Kulikov}, \bibinfo{person}{I.G.
  Chernykh}, \bibinfo{person}{A.V. Snytnikov}, \bibinfo{person}{B.M. Glinskiy},
  {and} \bibinfo{person}{A.V. Tutukov}.} \bibinfo{year}{2015}\natexlab{}.
\newblock \showarticletitle{{AstroPhi}: A code for complex simulation of the
  dynamics of astrophysical objects using hybrid supercomputers}.
\newblock \bibinfo{journal}{\emph{Computer Physics Communications}}
  \bibinfo{volume}{186}, \bibinfo{number}{Supplement C} (\bibinfo{year}{2015}),
  \bibinfo{pages}{71 -- 80}.
\newblock
\showISSN{0010-4655}
\urldef\tempurl%
\url{https://doi.org/10.1016/j.cpc.2014.09.004}
\showDOI{\tempurl}


\bibitem[\protect\citeauthoryear{McCalpin}{McCalpin}{1995}]%
        {McCalpin1995}
\bibfield{author}{\bibinfo{person}{John~D. McCalpin}.}
  \bibinfo{year}{1995}\natexlab{}.
\newblock \showarticletitle{Memory Bandwidth and Machine Balance in Current
  High Performance Computers}.
\newblock \bibinfo{journal}{\emph{IEEE Computer Society Technical Committee on
  Computer Architecture (TCCA) Newsletter}} (\bibinfo{date}{Dec.}
  \bibinfo{year}{1995}), \bibinfo{pages}{19--25}.
\newblock


\bibitem[\protect\citeauthoryear{Plimpton}{Plimpton}{1995}]%
        {Plimpton1995}
\bibfield{author}{\bibinfo{person}{Steve Plimpton}.}
  \bibinfo{year}{1995}\natexlab{}.
\newblock \showarticletitle{{Fast Parallel Algorithms for Short-Range Molecular
  Dynamics}}.
\newblock \bibinfo{journal}{\emph{J. Comput. Phys.}} \bibinfo{volume}{117},
  \bibinfo{number}{1} (\bibinfo{year}{1995}), \bibinfo{pages}{1--19}.
\newblock
\showISSN{00219991}
\urldef\tempurl%
\url{https://doi.org/10.1006/jcph.1995.1039}
\showDOI{\tempurl}


\bibitem[\protect\citeauthoryear{Popov and Ustyugov}{Popov and
  Ustyugov}{2007}]%
        {Popov2007}
\bibfield{author}{\bibinfo{person}{M.~V. Popov} {and} \bibinfo{person}{S.~D.
  Ustyugov}.} \bibinfo{year}{2007}\natexlab{}.
\newblock \showarticletitle{Piecewise parabolic method on local stencil for
  gasdynamic simulations}.
\newblock \bibinfo{journal}{\emph{Computational Mathematics and Mathematical
  Physics}} \bibinfo{volume}{47}, \bibinfo{number}{12} (\bibinfo{date}{01 Dec}
  \bibinfo{year}{2007}), \bibinfo{pages}{1970--1989}.
\newblock
\showISSN{1555-6662}
\urldef\tempurl%
\url{https://doi.org/10.1134/S0965542507120081}
\showDOI{\tempurl}


\bibitem[\protect\citeauthoryear{Popov and Ustyugov}{Popov and
  Ustyugov}{2008}]%
        {Popov2008}
\bibfield{author}{\bibinfo{person}{M.~V. Popov} {and} \bibinfo{person}{S.~D.
  Ustyugov}.} \bibinfo{year}{2008}\natexlab{}.
\newblock \showarticletitle{Piecewise parabolic method on a local stencil for
  ideal magnetohydrodynamics}.
\newblock \bibinfo{journal}{\emph{Computational Mathematics and Mathematical
  Physics}} \bibinfo{volume}{48}, \bibinfo{number}{3} (\bibinfo{date}{01 Mar}
  \bibinfo{year}{2008}), \bibinfo{pages}{477--499}.
\newblock
\showISSN{1555-6662}
\urldef\tempurl%
\url{https://doi.org/10.1134/S0965542508030111}
\showDOI{\tempurl}


\bibitem[\protect\citeauthoryear{Shenoy}{Shenoy}{2018}]%
        {optaneannounce}
\bibfield{author}{\bibinfo{person}{Navin Shenoy}.}
  \bibinfo{year}{2018}\natexlab{}.
\newblock \bibinfo{title}{What Happens When Your PC Meets Intel Optane Memory?}
\newblock
\newblock
\urldef\tempurl%
\url{https://newsroom.intel.com/editorials/what-happens-pc-meets-intel-optane-memory/}
\showURL{%
\tempurl}


\bibitem[\protect\citeauthoryear{{Smelyanskiy}, {Sawaya}, and
  {Aspuru-Guzik}}{{Smelyanskiy} et~al\mbox{.}}{2016}]%
        {qs-arxiv}
\bibfield{author}{\bibinfo{person}{M. {Smelyanskiy}}, \bibinfo{person}{N.~P.~D.
  {Sawaya}}, {and} \bibinfo{person}{A. {Aspuru-Guzik}}.}
  \bibinfo{year}{2016}\natexlab{}.
\newblock \showarticletitle{{qHiPSTER}: The Quantum High Performance Software
  Testing Environment}.
\newblock \bibinfo{journal}{\emph{ArXiv e-prints}} (\bibinfo{date}{Jan.}
  \bibinfo{year}{2016}).
\newblock
\showeprint[arxiv]{quant-ph/1601.07195}
\urldef\tempurl%
\url{https://arxiv.org/abs/1601.07195v2}
\showURL{%
\tempurl}


\end{thebibliography}


\end{document}